\documentclass[a4paper,12pt]{article}

\usepackage{mathtools}
\usepackage{amssymb}
\usepackage{amsfonts}
\usepackage[footnotesize]{caption}
\usepackage[font=scriptsize]{subcaption}
\usepackage{color}
\usepackage{braket}
\usepackage{cite}
\usepackage{hyperref}
\usepackage{url}
\usepackage{multirow}
\usepackage{relsize}
\usepackage{fullpage}
\usepackage{makecell}
\usepackage{fullpage}

\newcommand{\overbar}[1]{\mkern 1.5mu\overline{\mkern-1.5mu#1\mkern-1.5mu}\mkern 1.5mu}
\newcommand{\pmatr}[1]{\begin{pmatrix} #1 \end{pmatrix}}
\newcommand{\simlt}{~\mbox{\smaller\(\lesssim\)}~}

\begin{document}

\begin{titlepage}
\begin{center}
{\bf\Large Leptogenesis in minimal predictive seesaw models} \\[12mm]
Fredrik~Bj\"{o}rkeroth$^{\star}$
\footnote{E-mail: {\tt f.bjorkeroth@soton.ac.uk}},
Francisco~J.~de~Anda$^{\dagger}$
\footnote{E-mail: \texttt{franciscojosedea@gmail.com}},
Ivo~de~Medeiros~Varzielas$^{\star}$
\footnote{E-mail: \texttt{ivo.de@soton.ac.uk}},
Stephen~F.~King$^{\star}$
\footnote{E-mail: \texttt{king@soton.ac.uk}}
\\[-2mm]

\end{center}
\vspace*{0.50cm}
\centerline{$^{\star}$ \it
School of Physics and Astronomy, University of Southampton,}
\centerline{\it
SO17 1BJ Southampton, United Kingdom }
\vspace*{0.2cm}
\centerline{$^{\dagger}$ \it
Departamento de F{\'i}sica, CUCEI, Universidad de Guadalajara, M{\'e}xico}
\vspace*{1.20cm}

\begin{abstract}
{\noindent
We estimate the Baryon Asymmetry of the Universe (BAU) arising from leptogenesis within a class of minimal predictive seesaw models involving two right-handed neutrinos and simple Yukawa structures with one texture zero. The two right-handed neutrinos are dominantly responsible for the ``atmospheric'' and ``solar'' neutrino masses with Yukawa couplings to $(\nu_e, \nu_{\mu}, \nu_{\tau})$ proportional to $(0,1,1)$ and $(1,n,n-2)$, respectively, where $n$ is a positive integer. The neutrino Yukawa matrix is therefore characterised by two proportionality constants with their relative phase providing a leptogenesis-PMNS link, enabling the lightest right-handed neutrino mass to be determined from neutrino data and the observed BAU. We discuss an $SU(5)$ SUSY GUT example, where $A_4$ vacuum alignment provides the required Yukawa structures with $n=3$, while a $\mathbb{Z}_9$ symmetry fixes the relatives phase to be a ninth root of unity. }
\end{abstract}
\end{titlepage}

\section{Introduction}

Despite the success of the Standard Models (SM) of Particle Physics and Cosmology, the origin of matter-antimatter asymmetry in the Universe remains as a puzzling but unexplained phenomenon. Sakharov discovered that CP violation is a necessary condition for explaining the matter-antimatter asymmetry of the Universe \cite{Sakharov:1967dj}, but it became clear that the observed quark CP violation is insufficient for this purpose \cite{Kuzmin:1985mm}, motivating new sources of CP violation beyond the SM. One example of such new physics is neutrino mass and mixing \cite{King:2013eh}.
Indeed, following the discovery of a sizeable leptonic reactor angle, it is possible that leptonic CP violation could be observed in the foreseeable future through neutrino oscillations and a first tentative hint for a value of the CP-violating phase $\delta_{\mathrm{CP}} \sim -\pi /2$ has also been reported in global fits \cite{Gonzalez-Garcia:2014bfa,Forero:2014bxa,Capozzi:2013csa}. However the mass squared ordering (normal or inverted), the scale (mass of the lightest neutrino) and nature (Dirac or Majorana) of neutrino mass so far all remain unknown.%
\footnote{The neutrino mass squared ordering and the smallness of the lightest neutrino mass are commonly referred to jointly as the ``mass hierarchy'', although really they are two separate questions.}

There are many possibilities for the origin of light neutrino masses $m_i$ and mixing angles $\theta_{ij}$. Perhaps the simplest and most elegant idea is the classical seesaw mechanism, in which the observed smallness of neutrino masses is due to the heaviness of right-handed Majorana neutrinos \cite{Minkowski:1977sc},
\begin{equation}
m^{\nu}=-m^DM^{-1}_{R}(m^D)^T,
\label{seesaw}
\end{equation}
where $m^{\nu}$ is the light effective left-handed Majorana neutrino mass matrix (i.e. the physical neutrino mass matrix), $m^D$ is the Dirac mass matrix (in LR convention) and $M_R$ is the (heavy) Majorana mass matrix. The seesaw mechanism also provides an attractive mechanism for understanding the matter-antimatter asymmetry of the Universe. The out-of-equilibrium decays of right-handed neutrinos in the early Universe, combined with CP violation of the Yukawa couplings, leads to a lepton asymmetry which can be subsequently converted into a baryon asymmetry via sphaleron processes, a mechanism dubbed ``leptogenesis'' \cite{Fukugita:1986hr}.

Thermal leptogenesis in particular is an attractive and minimal mechanism to generate the Baryon Asymmetry of the Universe (BAU) $n_B/n_{\gamma} = (6.10 \pm 0.04) \times 10^{-10}$, or, normalised to the entropy density, $Y_B = (0.87 \pm  0.01) \times 10^{-10}$ (see e.g. \cite{DiBari:2012fz} for reviews and \cite{Ade:2015xua} for more recent determinations of the error which is now at $0.6\%$). A lepton asymmetry is dynamically generated and then converted into a baryon asymmetry due to $(B + L)$-violating sphaleron interactions which exist in the SM and its minimal supersymmetric extension, the MSSM. Leptogenesis can be implemented within the seesaw model, consisting of the SM (MSSM) plus right-handed Majorana neutrinos (and their superpartners) with a hierarchical spectrum. In the simplest case, the lightest of the right-handed neutrinos are produced by thermal scattering, and subsequently decay out-of-equilibrium in a lepton number and CP-violating way, thus satisfying the Sakharov constraints.

Although the seesaw mechanism generally predicts Majorana neutrinos, it does not predict the ``mass hierarchy'', nor does it yield any understanding of lepton mixing. In order to overcome these deficiencies, the seesaw mechanism must be supplemented by other ingredients. The idea of  ``sequential dominance'' (SD)  \cite{King:1998jw,King:2002nf} is that one dominant right-handed neutrino $N^{\rm atm}_R$ of mass $M_{\rm atm}$ is mainly responsible for the heaviest atmospheric neutrino mass $m_3$, while a second subdominant right-handed neutrino $N^{\rm sol}_R$ of mass $M_{\rm sol}$ mainly gives the solar neutrino mass $m_2$. These models were the first examples in the literature of two right-handed neutrino models \cite{King:1998jw}. One may add a third right-handed neutrino $N^{\rm dec}_R$ of mass $M_{\rm dec}$ but since this is essentially decoupled from the seesaw mechanism, we shall ignore it here. Sequential dominance immediately predicts a normal neutrino mass hierarchy.

In order to obtain sharp predictions for lepton mixing angles, the relevant Yukawa coupling ratios need to be fixed, for example using vacuum alignment of family symmetry breaking flavons \cite{King:2013eh}. The first attempt to use vacuum alignment within an $SU(3)$ family symmetry to predict maximal atmospheric mixing ($\tan \theta_{23} \sim 1$) from equal dominant right-handed neutrino couplings $y^i_{\rm atm}=(0,a,a)$ was discussed in \cite{King:2001uz}. Subsequently, constrained sequential dominance (CSD) \cite{King:2005bj} was proposed to explain tri-bimaximal mixing with a zero reactor angle by using vacuum alignment to fix the subdominant ``solar'' right-handed neutrino couplings to $(\nu_e, \nu_{\mu}, \nu_{\tau})$ to also be equal up to a sign,
namely $y^i_{\rm sol}=(b,b,-b)$. Other variants of SD with this vacuum alignment were used to obtain near tri-bimaximal mixing in the context of $SO(10)$-inspired GUTs \cite{deMedeirosVarzielas:2005ax}.
Following the measurement of the reactor angle, other types of CSD have been proposed, with the dominant right-handed ``atmospheric'' couplings as above, while proposing alternative subdominant ``solar'' right-handed neutrino couplings as follows:
\begin{equation}
\setlength{\itemsep}{-1.5ex}
	{\rm CSD}(n): \qquad (y_{\rm atm})^T=(0,a,a), \qquad (y_{\rm sol})^T=(b,nb,(n-2)b),\hspace{10ex}
\label{sol}
\end{equation}
where $n$ is any positive integer. The original CSD corresponds to $n=1$ \cite{King:2005bj}, with $n=2$ proposed in \cite{Antusch:2011ic} and $n=3,4$ in \cite{King:2013iva,King:2013xba,King:2013hoa,Bjorkeroth:2015ora}. A general analysis for $n<10$ was recently performed in \cite{Bjorkeroth:2014vha} where the smallest $\chi^2$ values were shown to correspond to $n=3,4$.%
\footnote{The CSD($n$) models may be regarded as a special class of two right-handed neutrino models with one texture zero \cite{King:2013iva}.
Two right-handed neutrino models with two texture zeros are only consistent with data for the case of an inverted mass squared ordering \cite{Harigaya:2012bw}. See also \cite{Zhang:2015tea} for a recent study of texture zeros in models with two right-handed neutrinos.}

In the original form of CSD, the columns of the Dirac mass matrix in the flavour basis were orthogonal to each other, $(y_{\rm atm})^{\dagger}y_{\rm sol}=0$, and consequently the CP asymmetries for cosmological leptogenesis vanished \cite{Antusch:2006cw,King:2006hn}. Following the subsequent observation that leptogenesis also vanished for a range of other family symmetry models \cite{Jenkins:2008rb}, this undesirable feature was eventually understood \cite{Choubey:2010vs} to be a general consequence of seesaw models with form dominance \cite{Chen:2009um} (i.e. in which the columns of the Dirac mass matrix in the flavour basis are proportional to the columns of the PMNS mixing matrix).%
\footnote{Although less minimal, a possibility to obtain non-vanishing leptogenesis is to have the interplay of more than one type of seesaw \cite{AristizabalSierra:2011ab}.}
However for CSD($n>1$), leptogenesis does not vanish since the columns of the Dirac mass matrix in the flavour basis are not orthogonal, $(y_{\rm atm})^{\dagger}y_{\rm sol}\neq 0$.

Interestingly, since the seesaw mechanism in CSD($n$) with two right-handed neutrinos only involves a single phase $\eta = \arg [a^2/b^2]$, both the leptogenesis asymmetries and the neutrino oscillation phase $\delta_{\mathrm{CP}}$ must necessarily originate from $\eta$, providing a direct link between the two CP violating phenomena in this class of models. Such a link can be traced back to the arguments in \cite{King:2002qh}.%
\footnote{The link between CP violation in flavour dependent leptogenesis and neutrino oscillation for models with sequential dominance was also observed in \cite{Antusch:2006cw}, although with only one leptogenesis phase the conclusions are identical to those obtained in the flavour independent or ``vanilla'' case \cite{King:2002qh}. Such a link was originally observed for the flavour independent case of two right-handed neutrino models with two texture zeros in \cite{Frampton:2002qc}, however such models are now phenomenologically excluded \cite{Harigaya:2012bw} for the case of a normal neutrino mass hierarchy considered here.}
The produced BAU $Y_B$ from leptogenesis in two right-handed neutrino models with CSD($n$) satisfies, following the arguments in \cite{Antusch:2006cw},
\begin{equation}\label{eq:BAU}
Y_B \propto \pm\sin \eta \: ,
\end{equation}
where the ``$+$'' sign applies to the case $M_{\rm atm}  \ll M_{\rm sol}$ and the ``$-$'' sign holds for the case $M_{\rm sol} \ll M_{\rm atm}$. Since the observed BAU $Y_B$ is positive, it follows that, for $M_{\rm atm}  \ll M_{\rm sol}$, we must have positive $\sin \eta$, while for $M_{\rm sol} \ll M_{\rm atm}$ we must have negative $\sin \eta$. From the analysis in \cite{Bjorkeroth:2014vha}, for CSD($n$) positive $\eta$ is associated with negative $\delta_{\mathrm{CP}}$ and {\it vice versa}. Although the global fits do not distinguish the sign of $\eta$, the present hint that $\delta_{\mathrm{CP}} \sim -\pi/2$ would require positive $\eta$, then in order to achieve positive $Y_B$ we require $M_{\rm atm}  \ll M_{\rm sol}$, corresponding to ``light sequential dominance'', as considered in the two right-handed neutrino analysis in \cite{Antusch:2011nz}. 

In this paper we estimate the baryon asymmetry arising from leptogenesis within CSD($n$) seesaw models with two right-handed neutrinos. In the flavour basis, for each $n$, the neutrino Yukawa matrix is therefore characterised by just two real proportionality constants plus one relative phase which controls both leptogenesis and the PMNS mixing matrix. The single phase appearing in the neutrino mass matrix is identified as the leptogenesis phase, providing a direct link between CP violation in neutrino physics and in cosmology. We use the observed BAU to constrain the mass spectrum of the two right-handed neutrinos within this class of models. As an example, we apply our results to a successful $A_4\times SU(5)$ SUSY GUT model based on CSD(3) with two right-handed neutrinos \cite{Bjorkeroth:2015ora}.

The layout of the remainder of this paper is as follows. In Section \ref{sec:neutrinos} we review thermal leptogenesis in seesaw models with two right-handed neutrinos, and apply it to CSD($n$). In Section \ref{sec:constrain} we show how low energy data constrains leptogenesis in these models, and derive bounds on the lightest right-handed neutrino mass. In Section \ref{sec:GUT} we review a SUSY GUT model that predicts CSD(3) mixing angles and reinterpret the bound imposed by BAU by expressing it in terms of the model's parameters. Finally we conclude in Section \ref{sec:Conclusions}.

\section{Leptogenesis formalism \label{sec:neutrinos}}
In a Supersymmetric (SUSY) model, the relevant terms in the superpotential giving neutrino masses are, in the diagonal charged lepton basis,
\begin{equation}
	W_\nu = y_{\mathrm{atm}}^i H L_i N_\mathrm{atm}^c 
	+ y_{\mathrm{sol}}^i H L_i N_\mathrm{sol}^c 
	+ M_{\rm atm} N_\mathrm{atm}^c N_\mathrm{atm}^c 
	+ M_{\rm sol} N_\mathrm{sol}^c N_\mathrm{sol}^c,
\label{eq:neutrinomassW}
\end{equation}
where $L_i$ are three families of lepton doublets, and the (CP conjugated) right-handed neutrinos $N_{\rm atm}^c$ and $N_{\rm sol}^c$ with real positive masses $M_{\rm atm}$ and $M_{\rm sol}$ do not mix. Assuming the CSD($n$) relations in Eq.~\ref{sol}, the Yukawa matrices and (charge conjugated) right-handed mass matrix in this basis are
\begin{equation}
	\lambda^{\nu} = \pmatr{0&b\\a&nb\\a&(n-2)b} ,\qquad\qquad 
	M^{c} = \pmatr{M_1 &0\\[2ex] 0&M_2},
	\label{seesawc}
\end{equation}
where we have written $M_1=M_{\rm atm}$ and $M_2=M_{\rm sol}$, with $M_1<M_2$, in anticipation of the result that the lighter right-handed neutrino is the dominant one. This is the basis used for the leptogenesis calculations, which we now review, following the notation and procedure in \cite{Antusch:2006cw}.

In the MSSM the regime where all flavours in the Boltzmann equations are to be treated separately corresponds to $(1+\tan^2 \beta)\times 10^5\ {\rm GeV}\simlt  M_1 \simlt (1+\tan^2 \beta)\times 10^9\ {\rm GeV}$. Assuming the flavour-dependent treatment for seesaw models with SD \cite{Antusch:2006cw}, it will turn out that for the models of interest $M_1\sim (40-100)\times 10^9\ {\rm GeV}$. The results therefore appear to post-justify the flavour-dependent treatment only for $\tan \beta \gtrsim 10$. However as it turns out, our results for the three flavour case are almost identical to those for the flavour independent case. The reason is that the washout factors for the $\mu$ and $\tau$ flavours turn out to be equal, $\eta_{1,\mu} =\eta_{1,\tau}$, while the asymmetry for the electron flavour is zero, $\epsilon_{1,e}=0$. 
In this case one may define an efficiency factor $\eta^{\rm ind}\equiv \eta_{1,\mu} =\eta_{1,\tau}$ and asymmetry $\epsilon_1 \equiv \sum_{\alpha} \epsilon_{1,\alpha }$ such that the BAU is proportional to $\eta^{\rm ind} \epsilon_1 $, as will become clear from the following results. The only difference between the flavour independent and dependent cases, in the considered models, is in the detailed solutions to the Boltzmann equations which involve differences in an $A$ matrix (defined below) which only appears logarithmically in determining the washouts. The main consequence of this is that the above condition $\tan \beta \gtrsim 10$ becomes relaxed, and our results are valid for any value of $\tan \beta$ to good approximation. However, for clarity, we shall perform our calculations using the full three flavour treatment.

Following  \cite{Antusch:2006cw}, the total BAU is obtained from the individual lepton flavour contributions:
\begin{equation}
	Y_B=\frac{10}{31}\sum_\alpha Y_{\Delta_\alpha},
	\label{eq:YB}
\end{equation}
which in turn are given by
\begin{equation}
	Y_{\Delta_\alpha} = \eta_{1,\alpha} [Y_{N_1} + Y_{\tilde{N}_1}] \epsilon_{1,\alpha},
	\label{eq:YDelta}
\end{equation}
where $\eta_{1,\alpha}$ are washout factors and $\epsilon_{1,\alpha}$ are the decay asymmetries. In the Boltzmann approximation for the MSSM:
\begin{equation}
	Y_{N_1}\approx Y_{\tilde{N}_1} \approx \frac{45}{\pi^4 g_*} , \qquad g_{*} =228.75.
	\label{eq:YN1}
\end{equation}

In the MSSM, the expression (per flavour index) for the asymmetries is
\begin{equation}
	\epsilon_{1,\alpha} = \frac{1}{8\pi} 
	\frac
		{\mathrm{Im}\left[(\lambda_\nu^\dagger)_{1\alpha}(\lambda_\nu^\dagger\lambda_\nu)_{12}(\lambda_\nu^T)_{2\alpha}\right]}
		{(\lambda_\nu^\dagger \lambda_\nu)_{11}} 
	g^{\mathrm{MSSM}}\left(\frac{M^2_2}{M^2_1}\right),
\end{equation}
where $\lambda_\nu$ is the neutrino Dirac Yukawa matrix in the basis when charged lepton and right-handed neutrino Yukawa matrices are diagonal, as in Eq.~\ref{seesawc}. Assuming $M_1\ll M_2$ we have:
\begin{equation}
	g^{\mathrm{MSSM}} \left(\frac{M^2_2}{M^2_1}\right) \approx -3 \frac{M_1}{M_2}.
\end{equation}

For the CSD($n$) neutrino Yukawa matrix $ \lambda_{\nu} $ in Eq.~\ref{seesawc}, the flavour dependent asymmetries are
\begin{align}
\begin{split}
	\epsilon_{1,e} 		&= 0 \\ 
	\epsilon_{1,\mu}	&= -\frac{3}{8\pi} \frac{M_1}{M_2} (n-1)n~\frac{\mathrm{Im}[a^{*2}b^2]}{|a|^2} \\ 
	\epsilon_{1,\tau}	&= -\frac{3}{8\pi} \frac{M_1}{M_2} (n-1) (n-2)~\frac{\mathrm{Im}[a^{*2}b^2] }{|a|^2}.
\end{split}
\label{eq:epsilonreduced}
\end{align}
Note that 
\begin{equation}
\epsilon_{1,\tau}=\left(\frac{n-2}{n}\right) \epsilon_{1,\mu}.
\end{equation}
We now define the phase $ \eta $ that is relevant for leptogenesis as
\begin{equation}
	\eta \equiv -\arg[a^{*2}b^2].
	\label{eq:eta1}
\end{equation}

Having established the factor $Y_{N_1}+Y_{\tilde{N}_1}$ and the $\epsilon_{1,\alpha}$ asymmetries, it remains to determine the (flavour-dependent) washout factors $ \eta_{1,\alpha} $. These arise from solutions to the Boltzmann equations given in \cite{Antusch:2006cw} wherein $ \eta_{1,\alpha} $ is plotted as a function of $ \log_{10}|A_{\alpha\alpha} K_\alpha| $, in terms of parameters $ K_\alpha $ and a numerical $ 3\times3 $ matrix $ A $, given below,
\begin{equation}
	A = \begin{pmatrix*}[r]
	-\frac{93}{110}& \frac{6}{55} & \frac{6}{55} \\[0.5ex] 
	\frac{3}{40} & -\frac{19}{30} & \frac{1}{30} \\[0.5ex] 
	\frac{3}{40} & \frac{1}{30} & -\frac{19}{30}
	\end{pmatrix*}.
	\label{eq:matrixA}
\end{equation}
The parameters $ K_\alpha $ are themselves functions of mass parameters $ \tilde{m}_{1,\alpha} $, such that
\begin{align}
\begin{split}
	K_\alpha &=\frac{\tilde{m}_{1,\alpha}}{m_{\mathrm{MSSM}}^*} , \qquad \quad \Big( K = \sum_{\alpha} K_\alpha \Big) \\[1ex]
	m_{\mathrm{MSSM}}^{*} &\approx (1.58\times 10^{-3}~\mathrm{ eV}) \sin^2 \beta \\
	\tilde{m}_{1,\alpha} &= (\lambda_\nu^\dagger)_{1\alpha} (\lambda_\nu)_{\alpha 1} \frac{v_u^2}{M_1}	.
\end{split}
\end{align}
With $ \lambda_\nu $ given in Eq.~\ref{seesawc}, and identifying $ v_u = v \sin \beta $, the mass parameters are
\begin{equation}
	\tilde{m}_{1,e}=0, \qquad \tilde{m}_{1,\mu} = \tilde{m}_{1,\tau} = |a|^2 \frac{v^2\sin^2\beta}{M_1}.
	\label{eq:mtilde}
\end{equation}
Because $\tilde{m}_{1,\mu}=\tilde{m}_{1,\tau}$ we also obtain $K_{\mu} = K_{\tau}$. From Eq.~\ref{eq:matrixA} we obtain $A_{\mu \mu} = A_{\tau \tau} = -19/30$. Thus we conclude that $\eta_{1,\mu} =\eta_{1,\tau}$. We note that the combination of parameters involved in $\tilde{m}_{1,\mu}$ is related to the effective neutrino masses which are fitted in \cite{Bjorkeroth:2015ora}, which are dependent on the choice of vacuum alignment parameter $ n $.

We may now return to the expression for the observed asymmetry $ Y_B $ as per Eqs.~\ref{eq:YB}-\ref{eq:YDelta}, where
\begin{equation}
	Y_B = \frac{10}{31} \sum_{\alpha} \eta_{1,\alpha} [Y_{N_1} + Y_{\tilde{N}_1}] \epsilon_{1,\alpha}.
\end{equation}
Inserting the approximations for $ Y_{N_1} $ and $ Y_{\tilde{N}_1} $ from Eq.~\ref{eq:YN1} and the asymmetries $ \epsilon_{1,\alpha} $ from Eq.~\ref{eq:epsilonreduced}, we get
\begin{equation}
\begin{split}
	Y_B &= \frac{10}{31} \left( \eta_{1,\mu} \left[2 \frac{45}{\pi^4 g_{*}}\right] \epsilon_{1,\mu} + \eta_{1,\tau} \left[2 \frac{45}{\pi^4 g_{*}}\right] \frac{n-2}{n} \epsilon_{1,\mu} \right) \\
	&= \frac{10}{31} \eta_{1,\mu} \left[2 \frac{45}{\pi^4 g_{*}}\right] \left(\frac{2n-2}{n}\right) \left(-\frac{3}{8\pi} \frac{M_1}{M_2} (n-1)n~\frac{\mathrm{Im}[a^{*2}b^2]}{|a|^2} \right).
\end{split}
\end{equation}
We may express this in terms of the phase $ \eta $ defined in Eq.~\ref{eq:eta1}, noting that $ \mathrm{Im}[a^{*2} b^2] / |a^2| = - |b|^2 \sin \eta $, arriving at
\begin{equation}
	Y_B = \frac{675}{31 \pi^5 g_{*}} \frac{M_1}{M_2} \eta_{1,\mu}~(n-1)^2 |b|^2 \sin \eta.
	\label{eq:baasy}
\end{equation}

In the next section we will show that this expression can be interpreted in terms of the mass parameters of the effective neutrino mass matrix $ m_a $ and $ m_b $, where the washout factor $ \eta_{1,\mu} $ is directly dependent on $ m_a $, while $ |b|^2/M_2 \propto m_b $. These mass parameters are subsequently dependent on $ n $. It becomes immediately obvious that $ n=1 $ gives $ Y_B = 0 $, i.e. that tri-bimaximal mixing cannot give a non-zero baryon asymmetry.

\section{Constraining leptogenesis with neutrino data \label{sec:constrain}}

In order to constrain leptogenesis in CSD($n$) we need to use information about low energy neutrino masses and mixing. These estimates are based on the seesaw mechanism. For the seesaw mechanism we use a different basis in which the Yukawa matrices $Y^{e}$, $Y^{\nu}$ are defined in a LR convention by
\begin{equation}
	\mathcal{L}^{LR} = -H^dY^e_{ij}\overline{L}^i_L e^j_R - H^u Y^{\nu}_{ij}
	\overline{L}^i_L  \nu^{i}_R + \mathrm{h.c.},
\label{seesawbasis}
\end{equation}
where $i,j=1,2,3$ label the three families of lepton doublets $L_i$, right-handed charged leptons $e^j_R$ and right-handed neutrinos $\nu_R^j$ below the GUT scale; $H^u, H^d$ are the Higgs doublets which develop VEVs $v_u,\,v_d$. Subscripts $L$ denote left-handed fermions and subscripts $R$ denote right-handed fermions.

Although it may appear to be confusing to use two different bases, it is valuable to do so because the seesaw mechanism is much more transparent in this basis. The light Majorana neutrino mass matrix $m^\nu$ is defined by \( \mathcal{L}^{LL}_\nu = -\tfrac{1}{2} m^\nu \overline{\nu}_{L} \nu^{c}_L + \mathrm{h.c.} \), while the heavy right-handed Majorana neutrino mass matrix $M_R$ is defined by \(\mathcal{L}^{RR}_\nu = -\tfrac{1}{2} M_{R} \overline{\nu^c}_{R} \nu_{R} + \mathrm{h.c.} \). With these conventions, using the seesaw basis in Eq.~\ref{seesawbasis}, integrating out the right-handed neutrinos leaves an effective left-handed neutrino Majorana mass matrix $ m^\nu $ expressed by the simple formula
\begin{equation}
	m^{\nu} = -v_u^2 Y^{\nu} M^{-1}_{R} Y^{\nu \mathrm{T}}.
\label{eq:seesaw}
\end{equation}
This is the reason for introducing the seesaw basis.

There is a simple dictionary between the seesaw basis just introduced and the SUSY basis in Eq.~\ref{eq:neutrinomassW}, as follows: $Y^{\nu}=(\lambda^{\nu})^*$, while $M_R=(M^c)^*=M^c$. Hence the CSD($n$) relations in Eq.~\ref{seesawc} become, in the seesaw basis,
\begin{equation}
	Y^{\nu} = \pmatr{0&b^*\\a^*&nb^*\\a^*&(n-2)b^*}, \qquad\quad 
	M_R = \pmatr{M_1 &0\\[2ex] 0&M_2}.
	\label{seesawy}
\end{equation}
The seesaw mechanism produces the effective neutrino mass matrix
\begin{equation}
	m^\nu = m_a \pmatr{0&0&0\\0&1&1\\0&1&1} 
			+ m_b e^{i\eta} \pmatr{1&n&(n-2)\\n&n^2&n(n-2)\\(n-2)&n(n-2)&(n-2)^2} ,
\label{eq:mnu}
\end{equation}
where $m_a = v_u^2 |a|^2 / M_1$ and $m_b = v_u^2 |b|^2 / M_2$ and we have multiplied throughout by an overall phase which we subsequently drop, keeping only the (physical) relative phase
\begin{equation}
\eta \equiv \arg [a^2/b^2].
\label{eta2}
\end{equation}
The definition of the phase $\eta$ in Eq.~\ref{eta2} is consistent with Eq.~\ref{eq:eta1}, providing the link between leptogenesis and low energy neutrino phenomenology. Clearly the phase $\eta$ plays a dual role, as both the high energy leptogenesis phase in Eq.~\ref{eq:baasy} and as the low energy neutrino mass matrix phase in Eq.~\ref{eq:mnu}. The sign of $ \eta $ has a low energy phenomenological significance, as the neutrino mass matrix phase in Eq.~\ref{eq:mnu} fixes the leptonic Dirac phase $ \delta_{\mathrm{CP}} $. Specifically, a positive $ \eta $ uniquely leads to negative $ \delta_{\mathrm{CP}} $, and \emph{vice versa}. As experimental data hints at $ \delta_{\mathrm{CP}} \sim -\pi/2 $, the \emph{a posteriori} preferred solution has positive $ \eta $. The sign of $ \eta $ also has high energy cosmological significance since the leptogenesis phase in Eq.~\ref{eq:baasy} controls the sign and magnitude of the BAU. For example a positive $ \eta $, together with the requirement that the BAU is positive, implies that the lightest right-handed neutrino should be $N_1^c=N_\mathrm{atm}^c $, while $N_2^c=N_\mathrm{sol}^c $ should be somewhat heavier. In the next section we discuss a SUSY GUT model with CSD(3) which satisfies this constraint and can lead to successful leptogenesis.

We devote the remainder of this section to the numerical CSD($n$) results for both neutrino phenomenology and leptogenesis. To do this, we recognise that the BAU $ Y_B $ given in Eq.~\ref{eq:baasy} depends implicitly on $ m_a $ through $ \eta_{1,\alpha} $ and explicitly on $ m_b $: recalling the definition $ m_b = |b|^2 v_u^2 /M_2 $, we may rewrite Eq.~\ref{eq:baasy} as 
\begin{equation}
	Y_B = \frac{675}{31 \pi^5 g_{*}} \frac{M_1 m_b}{v_u^2} \eta_{1,\mu}(n-1)^2 \sin \eta.
	\label{eq:baasy2}
\end{equation}
The best fit values of $ m_a $, $ m_b $ and $ \eta $ are found to vary with $ n $, and to be largely independent of one another. They have been calculated in \cite{Bjorkeroth:2015ora} and are reproduced in Table \ref{tab:csdn}, along with their predictions for the PMNS mixing angles, CP violating phase and neutrino masses. Furthermore, values of $ m_a $, $ m_b $ and $ \eta $ that may be characterised as providing ``good'' fits (or at least fits with $ \chi^2 $ close to the minimal value) lie comfortably within $ \pm 10\% $ of their respective best fit values. We are left with an expression for $ Y_B $ that is linear in $ M_1 $, multiplied by a numerical factor that ultimately depends only on $ n $. Taking into account the variability of the mass matrix parameters, we estimate that the numerical factor may also vary by up to $ \pm 10\% $ without significantly impacting the fits to neutrino masses and mixing angles. In terms of placing bounds on $ M_1 $, this far outweighs the current error on the experimental value for $ Y_B $, which is approximately $ \pm 0.6\% $. CSD(2) predicts a best fit with $ \eta = 0 $, while CSD($n$) with $ n>5 $ predict best fits with $ \eta = \pi $, both giving $ \sin \eta = 0 $, which implies a zero baryon asymmetry. Furthermore, these values of $ n $ give very poor fits to lepton data. As such, they will not be discussed further here.

\begin{table}[ht]
\renewcommand{\arraystretch}{1.2}
\centering
\footnotesize
\begin{tabular}{| c | c  c  c | c  c  c  c  c  c || c |}
\hline
\rule{0pt}{4ex} $n$ 	& \makecell{$m_a$ \\ {\scriptsize (meV)}} & \makecell{$m_b$ \\ {\scriptsize (meV)}} & 
\makecell{$\eta$  \\ {\scriptsize (rad)}}  	& \makecell{$\theta_{12}$ \\ {\scriptsize ($^{\circ}$)}} & \makecell{$\theta_{13}$ \\ {\scriptsize ($^{\circ}$)}}  & \makecell{$\theta_{23}$ \\ {\scriptsize ($^{\circ}$)}} & \makecell{$\delta_{\mathrm{CP}}$ \\ {\scriptsize ($^{\circ}$)}} & \makecell{$m_2$ \\ {\scriptsize (meV)}} & \makecell{$m_3$ \\ {\scriptsize (meV)}} & $\chi^2$  \\ [2ex] \hline 
3 	& 27.3		& 2.62		& 2.17		& 34.4		& 8.39		& 44.5		& -92.2		& 8.69		& 49.5		& 3.98		\\ 
4 	& 36.6		& 1.95		& 2.63		& 34.3		& 8.72		& 38.4		& -120		& 8.61		& 49.8		& 8.82		\\ 
5 	& 45.9		& 1.55		& 2.88		& 34.2		& 9.03		& 34.4		& -142		& 8.53		& 50.0		& 33.8		\\ 
\hline
\end{tabular}
\caption{
Table of best fit parameters for two right-handed neutrino CSD($n$) model for $ 3 \leq n \leq 5 $. For comparison CSD(2) (not shown) has $\chi^2=95.1$. Note that we have fixed $\eta$ to be positive corresponding to negative $\delta_{\mathrm{CP}}$. We have not displayed the Majorana phases which are also predicted but practically unobservable since $m_1=0$ in this class of two right-handed neutrino models.  Angles refer to the PDG standard parametrisation \cite{pdg}.}
\label{tab:csdn}
\end{table}

Note that from Eq.~\ref{eq:mtilde}, we have  $\tilde{m}_{1,\mu} = \tilde{m}_{1,\tau} = m_a $, whose best fit values for each CSD($n$) are given in Table \ref{tab:csdn}. This enables us to estimate $\log_{10}(A_{\mu \mu} K_{\mu}) = \log_{10}(A_{\tau \tau} K_{\tau})$, from the results in Eqs.~\ref{eq:matrixA}-\ref{eq:mtilde}, with which we obtain the washout factors from the numerical solutions to the Boltzmann equations given in \cite{Antusch:2006cw}. Hence, for $ n = (3, 4, 5) $, we obtain the corresponding washout factors $ \eta_{1,\mu} = (0.0236, 0.0166, 0.0126) $. Inserting numerical values also for $ m_b $ and $ \eta $ from Table \ref{tab:csdn} into Eq.~\ref{eq:baasy2}, we arrive at the following predictions%
\footnote{We have used $\sin\beta\approx 1$ which is a good  approximation for $\tan\beta>3$.}:
\begin{equation}
\begin{aligned}
	\mathrm{CSD(3):} 
	& \quad Y_B \sim 2.2 \times 10^{-11} \left[\frac{M_1}{10^{10} ~\mathrm{GeV}}\right] 
	&\Rightarrow& \quad M_1 \sim 4.0 \times 10^{10} ~\mathrm{ GeV} \hspace{4ex} \\[1ex]
	\mathrm{CSD(4):} 
	& \quad Y_B \sim 1.5 \times 10^{-11}\left[\frac{M_1}{10^{10} ~\mathrm{GeV}}\right]
	&\Rightarrow& \quad M_1 \sim 5.8 \times 10^{10} ~\mathrm{ GeV} \\[1ex]
	\mathrm{CSD(5):} 
	& \quad Y_B \sim 0.86 \times 10^{-11}\left[\frac{M_1}{10^{10} ~\mathrm{GeV}}\right]
	&\Rightarrow& \quad M_1 \sim 10 \times 10^{10} ~\mathrm{ GeV}
\end{aligned}
\end{equation}
With $M_1$ fixed in each case, $|a|$ may be calculated to be of order $10^{-3}$ using $m_a = v_u^2 |a|^2 / M_1$, since $m_a$ is known. On the other hand only the combination $m_b = v_u^2 |b|^2 / M_2$ is fixed by neutrino data and the separate parameters $|b|$ and $M_2$ are not determined from leptogenesis.

\section{A SUSY GUT Example \label{sec:GUT}}

In this section we describe a fairly complete $A_4 \times SU(5)$ SUSY GUT model which implements CSD(3) with two right-handed neutrinos~\cite{Bjorkeroth:2015ora}. This model has the following virtues:
\begin{itemize}
\renewcommand{\labelitemi}{$\circ$}
\renewcommand{\itemsep}{-1ex}
\item It is fully renormalisable at the GUT scale, with an explicit $SU(5)$ breaking sector and a spontaneously broken CP symmetry.  
\item The MSSM is reproduced with R-parity emerging from a discrete $\mathbb{Z}_4^R$. 
\item Doublet-triplet splitting is achieved through the Missing Partner mechanism \cite{Masiero:1982fe}.
\item A $ \mu $ term is generated at the correct scale.
\item Proton decay is sufficiently suppressed.
\item It solves the strong CP problem through the Nelson-Barr mechanism \cite{Nelson:1983zb, Barr:1984qx}.
\item It explains the hierarchies in the quark sector, and successfully fits all of the quark masses, mixing angles and the CP phase, using only $\mathcal{O}(1)$ parameters.
\item It justifies the CSD(3) alignment which accurately predicts the leptonic mixing angles, as well as a normal neutrino mass hierarchy.
\item It involves two right-handed neutrinos with the lighter one dominantly responsible for the atmospheric neutrino mass.
\item A $ \mathbb{Z}_9 $ flavour symmetry fixes the phase $ \eta $ to be one of ninth roots of unity \cite{Ross:2004qn}.
\end{itemize}

Apart from $A_4\times  SU(5)$ the model also involves the discrete symmetries $\mathbb{Z}_9\times \mathbb{Z}_6\times \mathbb{Z}_4^R$. It is renormalisable at the GUT scale, but many effects, including most fermion masses, come from non-renormalisable terms that arise when heavy messenger fields are integrated out. Unwanted or potentially dangerous terms are forbidden by the symmetries and the prescribed messenger sector, including any terms that would generate proton decay or strong CP violation. Such terms may arise from Planck scale suppressed terms, but prove to be sufficiently small. Due to the completeness of the model, the field content is too big to be listed here, but the superfields relevant to leptogenesis are in Table \ref{ta:SMF}.
\begin{table}
\centering
\footnotesize
\begin{minipage}[b]{0.45\textwidth}
\centering
\begin{tabular}{| c | c c | c | c | c |}
\hline
\multirow{2}{*}{\rule{0pt}{4ex}Field}	& \multicolumn{5}{c |}{Representation} \\
\cline{2-6}
\rule{0pt}{3ex}			& $A_4$ & SU(5) & $\mathbb{Z}_9$ & $\mathbb{Z}_6$ & $\mathbb{Z}_4^R$ \\ [0.75ex]
\hline \hline
\rule{0pt}{3ex}%
$F$ 					& 3 & $\bar{5} $& 0 & 0 & 1 \\
$T_1$ 					& 1 & 10		& 5 & 0 & 1 \\
$T_2$ 					& 1 & 10		& 7 & 0 & 1 \\
$T_3$ 					& 1 & 10		& 0 & 0 & 1 \\
\rule{0pt}{3ex}%
$N_{\rm atm}^c$ 		& 1 & 1	 		& 7 & 3 & 1 \\
$N_{\rm sol}^c$ 		& 1 & 1 		& 8 & 3 & 1 \\
\rule{0pt}{3ex}%
$\Gamma$				& 1 & 1			& 0 & 3 & 1 \\[0.5ex]
\hline
\end{tabular}
\end{minipage}%
\qquad
\begin{minipage}[b]{0.45\textwidth}
\centering
\begin{tabular}{| c | c c | c | c | c |}
\hline
\multirow{2}{*}{\rule{0pt}{4ex}Field}	& \multicolumn{5}{c |}{Representation} \\
\cline{2-6}
\rule{0pt}{3ex}			& $A_4$ & SU(5) & $\mathbb{Z}_9$ & $\mathbb{Z}_6$ & $\mathbb{Z}_4^R$ \\ [0.75ex]
\hline \hline
\rule{0pt}{3ex}%
$H_5$					& 1 & 5			& 0 & 0 & 0 \\
$H_{\bar{5}}$			& 1 & $\bar{5}$	& 2 & 0 & 0 \\
$H_{45}$	 			& 1 & 45 		& 4 & 0 & 2 \\
$H_{\overline{45}}$ 	& 1 & $\overline{45}$ & 5 & 0 & 0 \\
\rule{0pt}{3ex}%
$\xi$ 					& 1 & 1			& 2 & 0 & 0 \\
$\theta_2$				& 1 & 1			& 1 & 4 & 0 \\
$\phi_{\rm atm}$		& 3 & 1 		& 3 & 1 & 0 \\
$\phi_{\rm sol}$		& 3 & 1 		& 2 & 1 & 0 \\[0.5ex]
\hline
\end{tabular}
\end{minipage}
\caption{Superfields containing SM fermions, the Higgses and relevant flavons.}
\label{ta:SMF}
\end{table}

The SM fermions are contained within superfields $F$ and $T_i$. The MSSM Higgs doublet $H_u$ originates from a combination of $H_{5}$ and $H_{45}$, and $H_d$ from a combination of $H_{\overbar{5}}$ and $H_{\overbar{45}}$. Having the Higgs doublets inside these different representations generates the correct relations between down-type quarks and charged leptons. Doublet-triplet splitting is achieved by the Missing Partner mechanism \cite{Masiero:1982fe}.

The field $ \xi $ which gains a VEV $v_\xi \sim 0.06 M_{\mathrm{GUT}} $ generates a hierarchical fermion mass structure in the up-type quark sector through terms like $ v_u T_i T_j (v_\xi/M)^{6-i-j}$, where $v_u$ is the VEV of $H_u$. It also partially contributes to the mass hierarchy for down-type quarks and charged leptons and provides the mass scales for the right-handed neutrinos as discussed later. It further produces a highly suppressed $ \mu $ term $ \sim (v_\xi/M)^8 M_{\mathrm{GUT}} $.

The up-type Yukawa matrix $ Y^u $ is highly nondiagonal while the down-type and charged lepton Yukawa matrices $ Y^d \sim Y^e $, derived from terms like $ F \phi T H $, are nearly diagonal. A small off-diagonal element in $ Y^{d/e} $ sources the quark CP phase. Even though the charged lepton Yukawa matrix is not diagonal, this induces negligible corrections (of $\mathcal{O}(1\%)$) to the CSD(3) alignment.

The relevant terms in the superpotential giving neutrino masses are thus
\begin{equation}
	W_\nu = y_1 H_{5} F\frac{\phi_\mathrm{atm}}{\braket{\theta_2}} N_\mathrm{atm}^c 
	+ y_2 H_{5} F\frac{\phi_\mathrm{sol}}{\braket{\theta_2}} N_\mathrm{sol}^c 
	+ y_3 \frac{\xi^2}{M_\Gamma} N_\mathrm{atm}^c N_\mathrm{atm}^c 
	+ y_4 \xi N_\mathrm{sol}^c N_\mathrm{sol}^c,
\label{eq:neutrinomassWmodel}
\end{equation}
where the $y_i$ are dimensionless couplings, expected to be $\mathcal{O}(1)$. The alignment of the flavon vacuum is fixed by the form of the superpotential, with $\phi_{\mathrm{atm}}$ and $ \phi_{\mathrm{sol}} $ gaining VEVs according to CSD(3):
\begin{equation}
	\braket{\phi_{\mathrm{atm}}} = v_{\mathrm{atm}} \pmatr{0\\1\\1} , \qquad\qquad \braket{\phi_{\mathrm{sol}}} = v_{\mathrm{sol}} \pmatr{1\\3\\1}.
\end{equation}
Note that the above superpotential resembles that in Eq.~\ref{eq:neutrinomassW}. By the same method as outlined previously, we can then obtain leptogenesis estimates. Indeed, we can identify the parameters $a$, $b$, $M_1$ and $M_2$ in terms of the fundamental parameters from the superpotential:
\begin{equation}
\begin{aligned}
		a &= y_1 \dfrac{ v_{\mathrm{atm}} }{\braket{\theta_2}}, & \qquad \qquad
		b &=  y_2 \dfrac{ v_{\mathrm{sol}}}{\braket{\theta_2}}, \\
		M_1 &= y_3 \dfrac{(v_\xi)^2}{M_\Gamma} , &
		M_2 &= y_4 v_\xi.
\end{aligned}
\label{eq:modelparams}
\end{equation}
For convenience we can also specify:
\begin{align}
	m_a = \frac{v_u^2 |a|^2 }{M_1} = \left| \dfrac{ y_1^2 v_u^2 v_{\mathrm{atm}}^2  M_\Gamma}{y_3 \braket{\theta_2}^2 v_\xi^2} \right| , \qquad
	m_b=  \frac{v_u^2 |b|^2 }{M_2} = \left| \dfrac{ y_2^2 v_u^2 v_{\mathrm{sol}}^2  }{y_4 \braket{\theta_2}^2 v_\xi} \right|.
\end{align}

The Abelian flavour symmetry $\mathbb{Z}_9$ fixes the leptogenesis phase $ \eta $ to be one of the ninth roots of unity, through a variant of the mechanism used in \cite{Ross:2004qn}. The particular choice $\eta = 2\pi/3$ can give the neutrino mixing angles with great accuracy. Furthermore, this phase corresponds to $\delta_{\mathrm{CP}} \approx -87^{\circ}$, consistent with hints from experimental data.

\begin{table}[ht]
\renewcommand{\arraystretch}{1.2}
\centering
\footnotesize
\begin{tabular}{| c | c  c  c | c  c  c  c  c  c |}
\hline
\rule{0pt}{4ex}%
$n$ 	& \makecell{$m_a$ \\ {\scriptsize (meV)}} & \makecell{$m_b$ \\ {\scriptsize (meV)}} & 
\makecell{$\eta$  \\ {\scriptsize (rad)}}  	& \makecell{$\theta_{12}$ \\ {\scriptsize ($^{\circ}$)}} & \makecell{$\theta_{13}$ \\ {\scriptsize ($^{\circ}$)}}  & \makecell{$\theta_{23}$ \\ {\scriptsize ($^{\circ}$)}} & \makecell{$\delta_{\mathrm{CP}}$ \\ {\scriptsize ($^{\circ}$)}} & \makecell{$m_2$ \\ {\scriptsize (meV)}} & \makecell{$m_3$ \\ {\scriptsize (meV)}} \\ [2ex] \hline 
\rule{0pt}{4ex}%
3 	& 26.57		& 2.684		& $ \dfrac{2\pi}{3} $	& 34.3		& 8.67		& 45.8		& -86.7		& 8.59		& 49.8 \\[1.7ex]
\hline
\end{tabular}
\caption{Best fit parameters and predictions for an $ A_4 \times SU(5) $ SUSY GUT with CSD(3) and a fixed phase $ \eta = 2\pi/3 $, as described in \cite{Bjorkeroth:2015ora}. Angles refer to the parametrisation in \cite{pdg}.}
\label{tab:model}
\end{table}

The relevant best fit parameters from our model are given in Table \ref{tab:model}, along with the model predictions for the leptonic mixing angles and neutrino masses, for $ \tan \beta = 5 $. From Eq.~\ref{eq:mtilde}, we see that $\tilde{m}_{1,\mu} = \tilde{m}_{1,\tau} = m_a = 26.57 ~\mathrm{meV} $. This gives us $\log_{10}(A_{\mu \mu} K_{\mu}) = \log_{10}(A_{\tau \tau} K_{\tau}) = 1.027$, with which we obtain the washout factors from the numerical solutions to the Boltzmann equations given in \cite{Antusch:2006cw},
\begin{equation}
	\eta_{1,\mu } = \eta_{1,\tau } \approx 0.0236.
\end{equation} 
The decay asymmetries given in Eq.~\ref{eq:epsilonreduced} are calculated:
\begin{equation}
\begin{split}
	\epsilon_{1,\mu} &= 3\epsilon_{1,\tau}=\frac{9}{4\pi}\frac{M_1 m_b}{v^2\sin^2\beta}\sin \eta \\
	&\approx 6.01\times 10^{-7}\left[\frac{M_1}{10^{10}\ {\rm GeV}}\right].
\end{split}
\end{equation}

Using the above estimates, we may obtain the BAU for this model from Eq.~\ref{eq:baasy2}:
\begin{equation}
	Y_B \approx 2.2 \times 10^{-11} \left[\frac{M_1}{10^{10} ~\mathrm{GeV}} \right].
\end{equation}
Comparison with the experimental value of $ Y_B $ thus fixes the lightest right-handed neutrino mass:
\begin{equation}
	M_1 \approx 3.9 \times 10^{10} ~\mathrm{GeV}.
\end{equation}

As shown in Eq.~\ref{eq:modelparams}, in this model the right-handed neutrino mass is $M_1=y_3 (v_\xi)^2/M_\Gamma$, where $M_\Gamma$ is the renormalizable mass of the messenger $\Gamma$  that allows this term and is expected to be $M_\Gamma \sim M_P$. This fixes the arbitrary dimensionless constant to be $y_3\sim 0.3$, hence the BAU is achieved without extra tuning of parameters. Fixing the mass $M_1$ also fixes the parameter $a$ in the Yukawa matrix to be $a \approx 0.006$, defined in our model by Eq.~\ref{eq:modelparams}. 

\section{Conclusions \label{sec:Conclusions}}

We have considered leptogenesis in minimal predictive seesaw models involving two right-handed neutrinos,
with a particularly simple Yukawa structure which is capable of reproducing the low energy neutrino data with very few parameters.
Within this class of models the two right-handed neutrinos are dominantly responsible for the ``atmospheric'' and ``solar'' neutrino masses with 
Yukawa couplings to $(\nu_e, \nu_{\mu}, \nu_{\tau})$ proportional to $(0,1,1)$ and $(1,n,n-2)$, respectively, where $n$ is a positive integer.
In this class of seesaw models, called CSD($n$), the neutrino Yukawa matrix only involves two complex 
proportionality constants $a,b$ which, together with the real and positive right-handed neutrino masses $M_{1,2}$, 
control all of neutrino physics. To be precise, in the flavour basis, the two neutrino mass squared differences as well as
the entire PMNS matrix are controlled by the magnitudes of $a,b$, the two right-handed neutrino masses $M_{1,2}$
and the relative phase $\eta = \arg [a^2/b^2]$. This is the only phase in the neutrino mass matrix and is also the only phase 
relevant for leptogenesis, so it may be regarded as the ``leptogenesis phase''.

The entire PMNS matrix, including all the lepton mixing angles as well as all low energy CP violation, is sensitive to the 
magnitude and sign of the ``leptogenesis phase'' $\eta$, which therefore controls both low energy CP violation and the BAU 
via leptogenesis, providing a direct link between these two phenomena. 
A simple consequence is that CP violation must be present in neutrino oscillations in order to achieve successful leptogenesis.
Moreover there is a correlation between the sign of the BAU and the sign of CP violation which may be observed in neutrino oscillations.
To be precise, if the dominant right-handed neutrino mainly responsible for the atmospheric neutrino mass is the lightest one, 
then positive BAU implies $ \sin \eta > 0$ which further implies $ \sin \delta_{\mathrm{CP}} < 0$.

Within this class of seesaw models we have performed a quantitative study of flavoured leptogenesis arising from the decay of 
the lighter right-handed neutrino, using the magnitude of the observed BAU to provide information about its mass $M_1$. 
We expressed our results in terms of $n$ and recover previously known results, such 
that leptogenesis can not take place for $n=1$ (corresponding to tri-bimaximal mixing). For other values of $n$ we have as a 
general conclusion that the electron flavour doesn't contribute (having a zero asymmetry), with the asymmetry parameters 
having a ratio that is a simple function of $n$: $\epsilon_{1,\mu}/\epsilon_{1,\tau}=n/(n-2)$. Additionally, the muon and tau 
flavour share the same wash-out parameter $\eta_{1,\mu}=\eta_{1,\tau}$.

The nuts and bolts of our procedure is to insert the best-fit values of each CSD($n$) model's input parameters (including the phase $\eta$) 
into the prediction for $Y_B$, using previously obtained results of fits to the leptonic mixing angles and neutrino squared mass differences. 
This allows us to express $Y_B$ with only the mass $M_1$ as a free parameter. In this way we were able to 
predict $M_1$ from the combination of observational values of $Y_B$, neutrino squared mass differences and leptonic mixing angles.

We then considered an example of an $SU(5)$ SUSY GUT model with many attractive features incorporating CSD(3) which predicts a fixed value 
$\eta = 2 \pi/3$ (which is very close to the best-fit value for CSD(3)). In this model, the prediction of $M_1$ originating from $Y_B$ is 
expressed in terms of fundamental parameters of the model, with the correct $Y_B$ obtained with natural mass scales (Planck mass) and 
dimensionless parameters that are order unity. A notable feature of the model is that it predicts that the lighter right-handed neutrino 
is the dominant one responsible for both the atmospheric neutrino mass and leptogenesis, leading to the correct sign of the BAU with the
prediction $\delta_{\mathrm{CP}} \approx -87^{\circ}$, consistent with hints from experimental data.

We conclude that the minimal predictive seesaw models based on CSD($n$) are great examples of models connecting low and high energy 
CP violation and where leptogenesis works very well to obtain the BAU. The models which fit low-energy data well, namely $n=3,4$, 
lead to successful leptogenesis with masses $M_1$ on the order of $5 \times 10^{10}$ GeV. The approach is consistent with SUSY GUTs 
of flavour as exemplified by the $A_4\times SU(5)$ model which naturally realises the assumed Yukawa structures and mass scales 
in terms of more fundamental parameters.

\section*{Acknowledgements}

This project has received funding from the European Union's Seventh Framework Programme for research, technological development and demonstration under grant agreement no PIEF-GA-2012-327195 SIFT.
The authors also acknowledge partial support from the European Union FP7 ITN-INVISIBLES (Marie Curie Actions, PITN- GA-2011- 289442) and CONACyT.

\end{document}